**Plasma-enhanced chemical vapor deposition of amorphous Si on graphene**

G. Lupina,[1,1)] C. Strobel,[2] J. Dabrowski,[1] G. Lippert,[1] J. Kitzmann,[1] H.M. Krause,[1] Ch. Wenger,[1] M. Lukosius,[1] A. Wolff,[1] M. Albert,[2] J.W. Bartha[2]

[1]IHP, Leibniz-Institut für innovative Mikroelektronik, Im Technologiepark 25, 15236 Frankfurt (Oder), Germany
[2]Institut für Halbleiter- und Mikrosystemtechnik, Technische Universität Dresden, 01062 Dresden, Germany

Plasma-enhanced chemical vapor deposition of thin a-Si:H layers on transferred large area graphene is investigated. Radio frequency (RF, 13.56 MHz) and very high frequency (VHF, 140 MHz) plasma processes are compared. Both methods provide conformal coating of graphene with Si layers as thin as 20 nm without any additional seed layer. The RF plasma process results in amorphization of the graphene layer. In contrast, the VHF process keeps the high crystalline quality of the graphene layer almost intact. Correlation analysis of Raman 2D and G band positions indicates that Si deposition induces reduction of the initial doping in graphene and an increase of compressive strain. Upon rapid thermal annealing the amorphous Si layer undergoes dehydrogenation and transformation into a polycrystalline film whereby a high crystalline quality of graphene is preserved.

Ability to deposit uniform thin layers of insulators and semiconductors on graphene is of crucial importance for many envisioned applications of this material in advanced electronic and photonic devices.[1] Si-graphene junctions are particularly interesting for graphene-based photodetectors, sensors, and high-frequency vertical heterojunction transistors.[2,3] Deposition of materials on graphene is generally considered very challenging due to the hydrophobic nature of graphene rendering conventional chemical vapor deposition (CVD) and atomic layer deposition (ALD) methods ineffective in the absence of additional seeding layer.[1] For example, we recently reported a seed-free growth of HfO$_2$ by CVD on transferred large area graphene[4] and concluded that although thick HfO$_2$ layers (>30 nm) are closed and show good electrical performance, thinner films are not homogenous and show a large

---
[1)] Author to whom correspondence should be addressed. Electronic mail: lupina@ihp-microelectronics.com

number of pinholes. On the contrary, evaporation (e.g. by e-beam) of materials onto CVD graphene results in a much more homogeneous layer even in the low thickness regime (few nm).[5,6] However, methods such as e-beam evaporation are often not compatible with large scale semiconductor device manufacturing. In contrast, both CVD and plasma enhanced CVD (PECVD) are widely accepted manufacturing methods. PECVD is particularly interesting for applications requiring low thermal budgets such as the back end of line (BEOL) semiconductor device fabrication (<~500°C). Integration of graphene in the BEOL-regime[7] naturally favors PECVD methods witch enable low deposition temperatures. However, high energy ion bombardment related to plasma exposure readily correlates with worsening of material properties.[8,9] High energy ion exposure can easily induce defects also in the crystalline lattice of graphene.[10-12] Such defects result in a dramatic degradation of graphene electrical properties seriously limiting its practical applications. Heintze and Zedlitz have shown that ion energy in plasmas decreases with increasing frequency up to at least 180 MHz.[13] For high plasma excitation frequencies plasma resistance and sheath width is decreased. Therefore voltage needed to maintain the plasma is reduced which also leads to lower plasma potentials. As the maximum ion energy in the plasma primarily scales with voltage and plasma potential, lower ion energies are present in high-frequency plasmas. For example, investigations of frequency dependent energetic distribution of ions impinging on the substrate in plasma enhanced deposition processes demonstrated that increasing the plasma excitation frequency to values above 120 MHz reduces the energy of ions to below 25 eV.[13] Further increase of the excitation energy results in further ion energy reduction. On the other hand, recent systematic studies of damage induced in graphene by ion bombardment indicate a strong correlation between the energy of impinging ions and the intensity of the defect-related Raman D-band.[14] In particular it was shown that the number of generated defects rapidly decreases when the kinetic ion energy is reduced below 25 eV.[14] Using plasma excitation frequencies larger than 100 MHz is a unique feature in the field of PECVD methods.[15] Beneficial effects of using 140 MHz PECVD compared to lower frequencies were already demonstrated by Leszczynska et al. for thin-film silicon solar cells.[16] Here, we show that in contrast to the conventional radio frequency (RF, 13.56 MHz) PECVD a very high frequency (VHF, 140 MHz) PECVD can be used to cover CVD graphene with thin a-Si:H layers very softly without changing the properties of the underlying graphene significantly. Direct comparison between the RF and VHF methods demonstrates a decisive advantage of the VHF plasma: seed layer-free conformal coverage of graphene at low temperatures without significant degradation of its crystalline quality. We associate this striking difference with reduced ion energy in VHF plasmas[13] and resulting significantly lower amount of bombardment-related defects in the hexagonal network of $sp^2$-hybridized carbon.

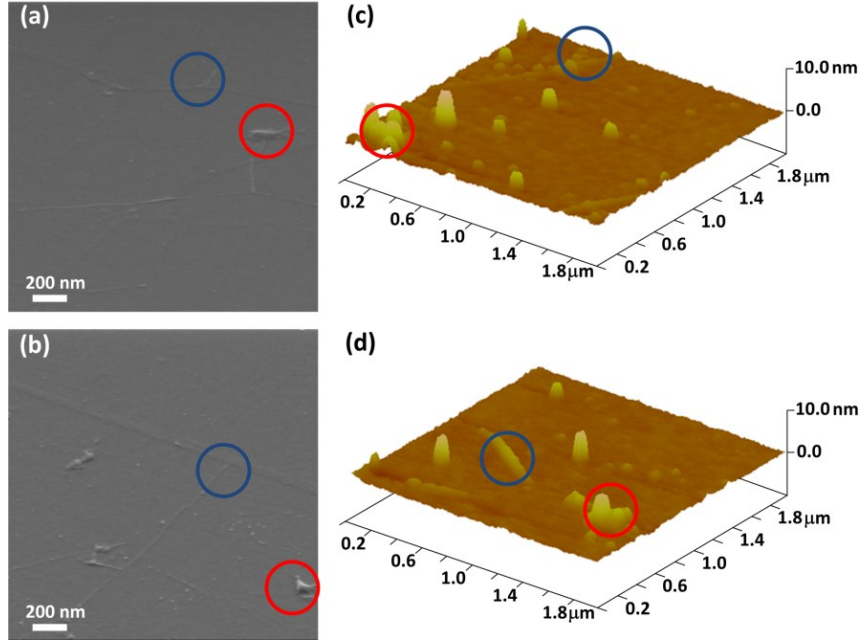

FIG. 1. SEM images taken after deposition of 20 nm of amorphous Si on Gr by VHF (a) and RF (b) PECVD. Panels (c) and (d) show corresponding AFM images obtained on VHF and RF layers, respectively. RMS roughness extracted from the AFM measurements (on area 1x1 um$^2$) is 0.67 nm and 0.61 nm for VHF and RF method, respectively. Blue circles mark graphene wrinkles and folds. Red circles mark residual backside graphene and other transfer related residuals.

Commercially available graphene on Cu foils was transferred onto 100 nm SiO$_2$ substrates using wet etch transfer technique.[17,18] All graphene samples (1cm x 1cm) for both RF PECVD and VHF PECVD were cut from the same piece of 5cm x 5cm-large graphene/Cu foil to ensure a fair comparison between both methods. Before Si deposition samples were annealed in UHV at 400°C for 20 min to remove residual polymer used as a support in the transfer process. To deposit thin a-Si:H layers on transferred graphene we used 140 MHz excitation[15] to exploit the ion energy reduction potential of VHF PECVD as much as possible. For comparison, a-Si:H layers of the same thickness (20 nm and 30 nm) were deposited also by conventional RF PECVD. In the following, we will refer to the samples fabricated by both methods as RF and VHF layers. The VHF and RF a-Si:H layers were deposited at 200 °C using silane precursor under exactly the same experimental conditions except the different plasma excitation frequency. Due to different power coupling and losses between RF and VHF the deposition rate was leveled off for both frequencies rather than the input power. The deposition rate for both methods was kept low (10 nm/min) to minimize ion bombardment. Pressure in the growth chamber was kept at 0.4 mbar during deposition. After Si growth samples were characterized by Raman, secondary electron microscopy (SEM), and atomic force microscopy (AFM). The root means square roughness (rms) was extracted from AFM measurements on 1x1 um$^2$ areas of the sample. Raman measurements were carried out at room temperature using Renishaw InVia spectrometer equipped with a 50× objective and a 514 nm laser source. Rapid thermal annealing (RTA) treatment was performed in $N_2$ ambient for 1 min.

Figure 1 presents SEM and AFM images acquired after deposition of about 20 nm-thick a-Si layer onto transferred graphene using VHF and RF PECVD methods. According to these investigations both techniques provide conformal coverage of the graphene layer without pinholes in the a-Si layer and rms roughness of 0.67 nm and 0.61 nm for VHF and RF technique, respectively. Inhomogeneity visible in the SEM and AFM images is caused mainly by the presence of wrinkles/folds and transfer-related residues of the backside graphene and/or the polymer support.[18]

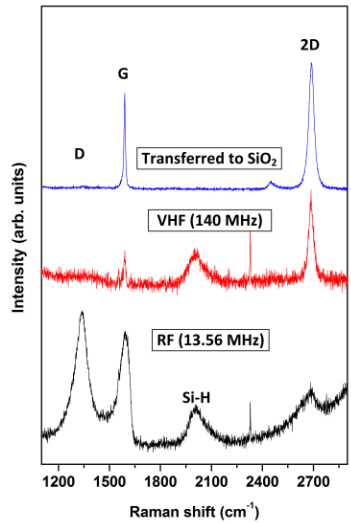

FIG. 2. Raman spectra showing G and 2D graphene bands for as transferred CVD graphene and after deposition of 20 nm of amorphous Si by RF (13.56 MHz) and VHF (140 MHz) PECVD. Worsening of the signal-to-noise ratio observed after Si deposition is associated with attenuation of the Raman signal from graphene by the Si overlayer.

Figure 2 shows an overview of Raman spectroscopy results obtained from graphene samples covered with Si layers using RF and VHF PECVD methods (with identical deposition parameters except the plasma frequency). Raman spectrum taken before Si deposition shows two strong peaks at ~ 1585 $cm^{-1}$ and 2680 $cm^{-1}$ which are assigned to the G and 2D bands, respectively. A negligible D band in this spectrum proves that a high quality graphene layer is obtained after transfer.[1] Subsequent deposition of a-Si layer results in several changes in the Raman spectra. Firstly, a broad peak centered at around 2000 $cm^{-1}$ becomes visible. This peak is associated with Si-H bond vibrations.[19] Secondly, appearance of a strong D-band (~1350 $cm^{-1}$) in the spectrum taken from the RF sample clearly indicates a large increase in the number of broken $sp^2$ carbon-carbon bonds in the graphene crystalline lattice. In contrast, after VHF PECVD process, D band is barely visible implying that a good crystalline quality of the graphene layer is preserved. For this reason the following detailed analysis of Raman spectra is performed for VHF samples only. Finally, significant shifts of the 2D and G peak positions and changes in their relative intensity are observed as a result of Si deposition. Such variations can be attributed to doping and strain in the graphene layer.[20,21]

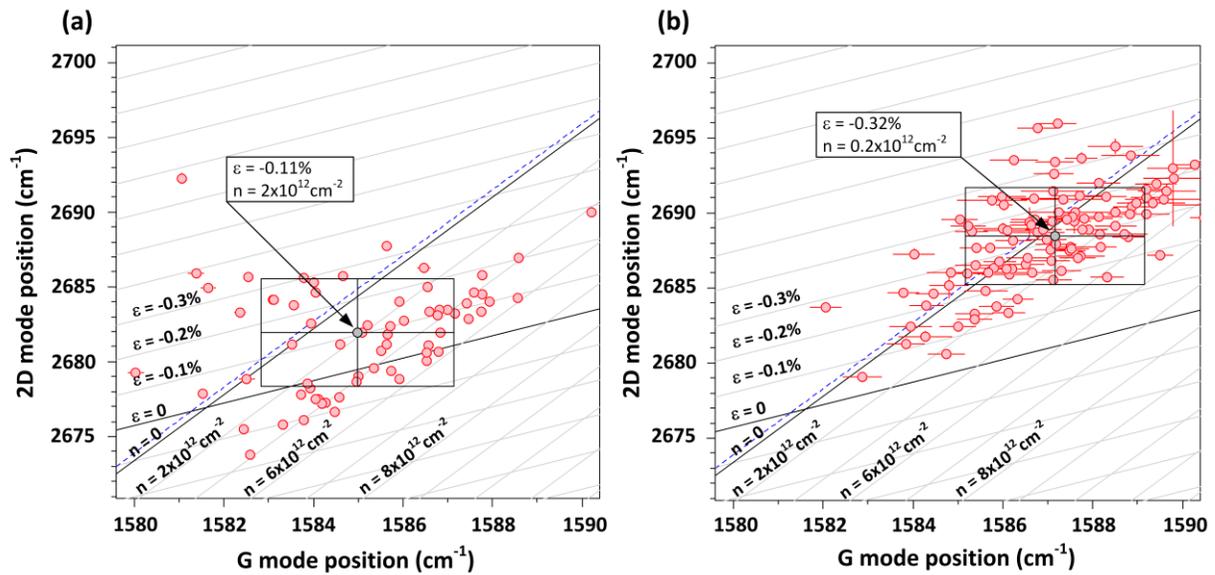

FIG. 3. Correlation between the G and 2D mode position before (a) and after (b) Si deposition by VHF PECVD. Black solid lines indicate the 2D and G peak position dependence for strained undoped and unstrained doped free-standing graphene (cf. Ref. 22). Before Si deposition a significant p-doping and a small compressive strain is detected. Si deposition results in a reduction of the initial p-type doping and an increase of compressive strain in the graphene sheet.

To distinguish between the strain and doping effects we use a correlation analysis of the G and 2D peak positions as proposed by Lee et al.[22] Raw data for this analysis were obtained from Raman maps acquired before and after Si deposition on area of 20x20 um$^2$ yielding 121 measurement points. A summary of this analysis is presented in Figure 3 a-b. There is a broad distribution in peak positions between individual measurements indicating presence of inhomogeneities in both graphene (wrinkles, multilayer islands, grain boundaries, etc.) as well as its environment (substrate roughness, local adsorbates, transfer-related impurities, etc.). Despite this broad distribution, clear trends can be recognized by considering the average positions of the investigated Raman peaks. Average value for the as-transferred graphene indicate a strong p-type doping (~$2x10^{12}$ cm$^{-2}$) which is likely due to adsorbed $O_2$ and/or moisture and consistent with previous reports.[23,24] As a result of Si deposition (Fig. 3b) there is a substantial increase in the average compressive strain in the graphene layer (from about -0.11% to -0.32%) and a reduction of the p-type doping (from about $2x10^{12}$cm$^{-2}$ to ~ $0.2x10^{12}$ cm$^{-2}$). The latter conclusion is corroborated by the results of 2D/G peak ratio analysis for both cases (Fig. 4a-b). 2D/G intensity ratio increases from its average value of about 1.2 before Si deposition to about 3.5 after the VHF PECVD process. This indicates that the Fermi level in graphene moves towards the charge neutrality point and the amount of the initial p-type doping is reduced.[25,26]

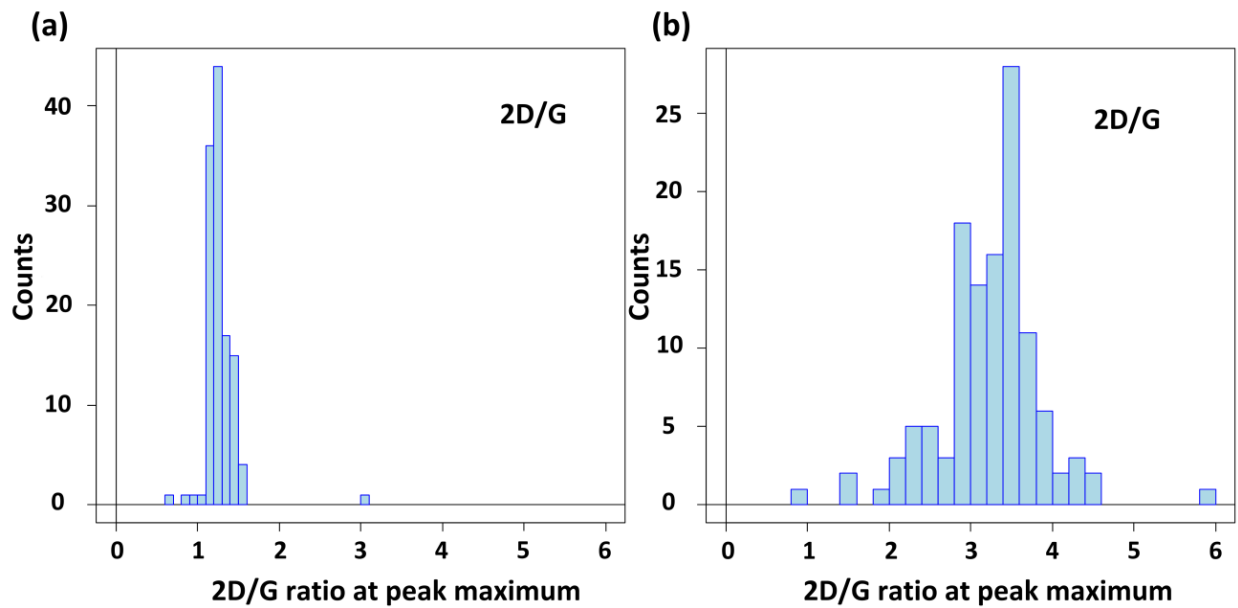

FIG. 4. Statistical distribution of the 2D to G peak intensity ratio before (a) and after (b) VHF PECVD Si deposition. Analysis is performed based on 20x20 um$^2$ Raman maps. 2D/G peak ratio clearly increases after Si deposition indicating shift of the Fermi level in graphene towards the charge neutrality point.

Statistical analysis of the D/G peak ratio (Fig. 5 a-b) reveals that for the majority of measurement points D/G is below 0.05 for as deposited graphene. After VHF Si deposition, most spectra show D/G ~ 0.2. This indicates that the high frequency plasma also induces a measurable degree of disorder in the graphene crystalline lattice. However, as the D/G is above 1 for the RF a-Si:H sample (cf. inset to Fig. 5b) and the corresponding 2D peak almost completely disappears in this case (cf. Fig.2) it is concluded that the VHF process induces much less defects and is thus more suitable for the preparation of Si layers on graphene.

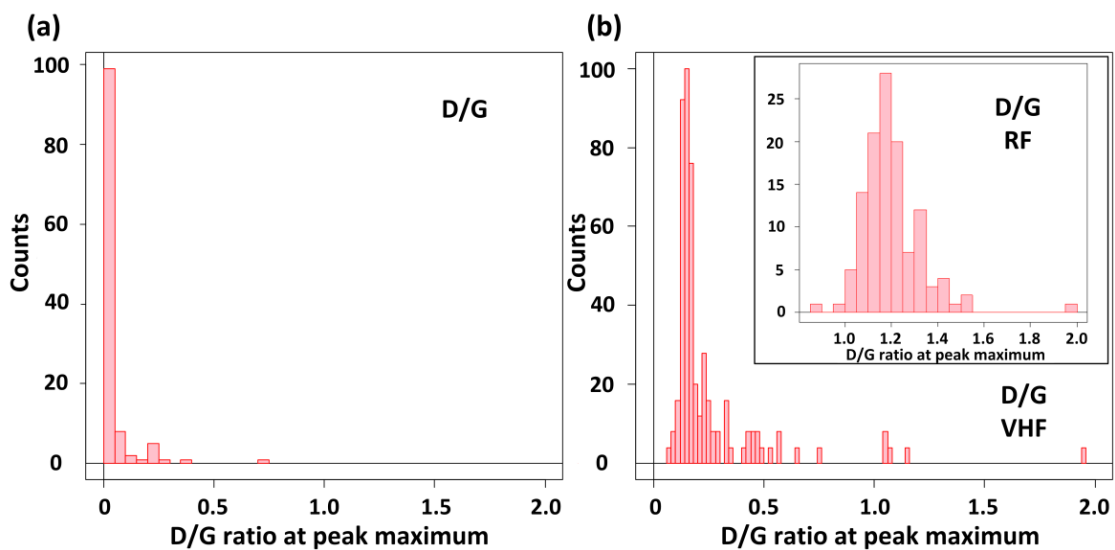

FIG. 5. Ratio of the D to G peak intensity before (a) and after (b) Si deposition.

Figure 6 summarizes Raman spectroscopy results obtained from a 20 nm thick VHF Si sample on graphene after several RTA steps in the temperature range of 500°C – 900°C. The broad band associated with the vibrations of Si-H bonds (Fig. 6a, around 2000 cm$^{-1}$) disappears after RTA at 500°C indicating dehydrogenation of the a-Si:H layer.[27] The corresponding spectra taken in the Si vibrational region (Fig. 6 b) prove that the Si layer remains amorphous after the 500°C RTA treatment step. As a result of RTA at 700°C, a broad band with a maximum at 480 cm$^{-1}$ assigned to a-Si vanishes and only a small shoulder (marked with arrow in Fig. 6b) with a local maximum at 501 cm$^{-1}$ is visible. This points out to a transformation of the amorphous Si layer on graphene into a polycrystalline film.[27] Broadening of the main Si band (FWHM at RT and 700°C of 3.8 cm$^{-1}$ and 5.4 cm$^{-1}$, respectively) and a small shift in the position of its maximum (521.0 cm$^{-1}$ at RT and 520.8 cm$^{-1}$ RTA at 700°C) indicate rise of an additional component close to the Si band from Si wafer substrate related to polycrystalline Si. Our Raman investigations provide no indications that the RTA treatment significantly degrades the quality of the underlying graphene film except the RTA treatment at 900°C which causes appearance of a weak D band (Fig. 6a). As a result of RTA a slight increase in surface roughness of the Si overlayer is observed. According to our AFM measurements, the rms roughness increased from 0.67 nm after Si deposition to 0.79 nm after RTA at 900°C. Further experiments focused on elucidating a possible impact of the PECVD and RTA processes on the electrical properties of graphene[28] are ongoing.

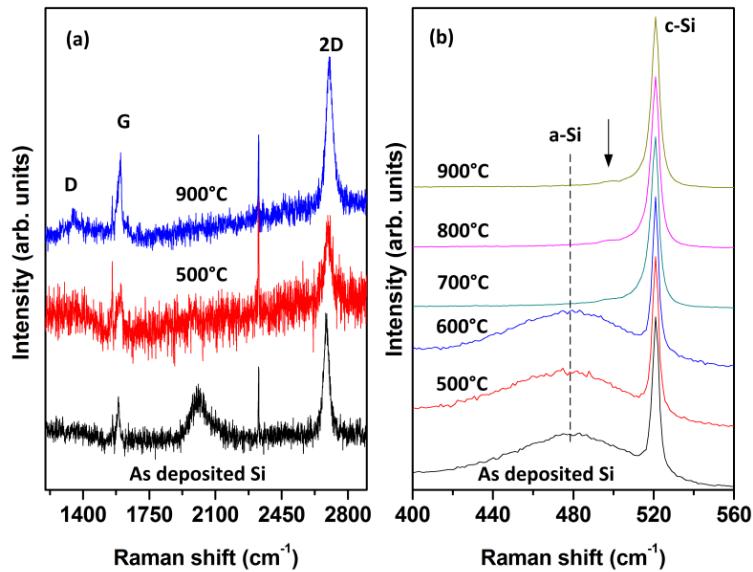

FIG. 6. Raman spectra of Gr/Si stack subjected to RTA treatment showing graphene bands (a) and Si band (b).

In summary, we investigated growth of a-Si:H on transferred CVD graphene using two different deposition methods: VHF (140 MHz) and RF (13.56 MHz) PECVD. Both methods provide conformal coverage of graphene with pinhole-free Si layers even in the low thickness regime (~20 nm). Correlation analysis of Raman 2D and G band positions indicates that Si deposition induces reduction of the initial p-type doping in graphene and increase of compressive strain. Upon rapid thermal annealing the amorphous Si layer undergoes dehydrogenation and transformation into a polycrystalline film whereby a high crystalline quality of graphene remains preserved. Our experiments reveal that while RF plasma results in an almost complete damage of the graphene crystalline lattice, the VHF plasma is very gentle inducing only a relatively small number of defects in graphene. We conclude that this is due to the reduced energy of ions which is expected for the VHF plasma. As a result, VHF PECVD appears as a very attractive scalable method enabling gentle seed layer-free formation of thin semiconductor layers on graphene which may open the way to device applications relying on Si-graphene junctions.

**Acknowledgments**

Financial support by the German Research Foundation in the framework of the Priority Program SPP FFlexCom (BA 2009/6-1 & WE 3594/5-1) and the European Commission through a STREP project (GRADE, No. 317839) is gratefully acknowledged.